\documentclass[letter]{ptptex}

\usepackage{bm}
\usepackage{graphicx}

\newcommand{\bx}{{\bm{x}}}
\newcommand{\br}{{\bm{r}}}
\newcommand{\bk}{{\bm{k}}}
\newcommand{\bp}{{\bm{p}}}

\title{
Distortion of the HBT images by meson clouds%
}

\author{
Koichi \textsc{Hattori}\footnote{hattori@hep1.c.u-tokyo.ac.jp} 
and T. \textsc{Matsui}\footnote{tmatsui@hep1.c.u-tokyo.ac.jp}%
}

\inst{
Institute of Physics, University of Tokyo\\
Komaba, Tokyo 153-8902, Japan
}

\abst{
We study the effects of mesonic final state interactions on the Hanbury Brown
and Twiss (HBT) intensity interferometry for mesons in ultra-relativistic heavy 
ion collisions.  
Modification of the one-body amplitude of emitted mesons while going 
through a cloud of other mesons is estimated in the semiclassical 
approximation with a mesonic optical potential which incorporates both
coherent forward scattering with other mesons and the absorption due to the 
incoherent scattering in the meson clouds.  
We show how these effects results in the distortion of the HBT images.
}

\begin{document}

\maketitle

Hanbury Brown and Twiss intensity interferometry, originally developed in
radio astronomy to measure stellar radii\cite{HBT56}, and later played
a seminal role in the development of quantum optics\cite{Glauber63}, 
has been applied to measure the source size of the secondary particles 
emitted in high energy heavy ion collisions.\cite{YK78,GKW79,Baym}
Recent analysis of the RHIC experiments has shown that the source 
shape determined from the data has a systematic deviation from 
the prediction of the hydrodynamical models \cite{STAR,PHENIX,
HT02,Lisa}: this has been called "RHIC HBT puzzle".
\footnote{
Very recent analyses \cite{imaging} of the RHIC data using elaborate "imaging technique" 
\cite{BD97} has claimed, however, to reproduce a source function constructed by an event
generator which incorporates a naive space-time picture for hadronization with extended 
resonance decay tail.}

The HBT formula for the correlation function usually used in data analysis
assumes: \cite{Chu}
1) random initial phases (incoherent source), 
2) factorization of two-point source function, 
3) neglection of all the interaction between two detected pions and the rest of the 
system after the emission. 
Each of these underlying assumptions may need to be checked in order to find a
resolution of the ''HBT puzzle''. 
In this work, we investigate the effects of the final state interactions on the 
correlation function, assuming the first two assumptions are valid.  

We focus on the effect of one-body interaction, namely the interacion 
between each of the observed pairs and rest of the system via mean 
field potential. 
The mutual interaction between the two detected pions is dominated at large 
separation, corresponding to small $q$, by long range Coulomb interaction, 
rather than strong interaction, and such effect has been incorporated by the 
well-known Gamow factor in the correlation function\cite{GKW79}. 
In the usual hydrodynamical modeling, the particle momenta are considered
to be frozen on the kinetic freeze-out surface.   
In the kinetic theoretical language, this implies that the collisions between two 
particles which maintains the system in local equilibrium becomes suddenly 
ineffective.  
However, there still remains interaction between each of the pair 
and the evaporating particles in the vicinity of the emission points.
We introduce mesonic optical potential to describe such effects: the real part 
of the potential describes a coherent forward scattering of the particle with the 
other mesons, while the imaginary absorptive part incorporates the effect of 
incoherent scattering with other individual mesons.  
We examine how extracted HBT images are distorted due to the modification
of the one-body amplitudes in the mesonic optical potential.

In the standard picture of ultrarelativistic nuclear collisions, the pion source
extends along the direction of the motion of colliding nuclei in an approximately boost 
invariant fashion\cite{Bjo83}.   We follow this picture, together with cylindrical
symmetry of the collision volume, and concentrate on the quantum evolution 
of a group of particles appearing in a small central rapidity bin 
$- \frac{1}{2} \Delta y < y < \frac{1}{2} \Delta y$ on the two-dimensional 
transverse plane in their center-of-mass frame.   

Two particle momentum correlation function of identical particle which we are
concerned is given by
\begin{eqnarray}
C(\bm{k}_1,\bm{k}_2)=\frac{P_2(\bm{k}_1,\bm{k}_2)}{P_1(\bm{k}_1)P_1(\bm{k}_2)}    
\end{eqnarray}
where $P_2(\bm{k}_1,\bm{k}_2)$ is the joint probability of detecting a pair of the same 
kind of pions with momenta $\bm{k}_1$ and $\bm{k}_2$,  and $P_1(\bm{k})$ is the probability 
for a detection of single pion.   
They may be given in terms of the asymptotic form of the matrix elements of the density 
matrix: 
$P_1(\bm{k}) =  \langle\bm{k}|\hat \rho(\infty)|\bm{k}\rangle$ ,
$P_2(\bm{k}_1,\bm{k}_2) = \langle \bm{k}_1\bm{k}_2 | \hat \rho(\infty)|\bm{k}_1\bm{k}_2
\rangle$. 
Here we adopt the Schr\"odinger picture in which the time evolution of the 
density matrix is given by $\hat \rho (t) = U (t) \hat \rho (0) U^\dagger (t)$ so that 
we may write 
\begin{eqnarray} 
P_1(\bm{k}) 
& = & \lim_{t \to \infty} \int d\bm{x}_1 d\bm{x}_2 \langle\bm{k}| U(t) |\bm{x}_1\rangle
\langle \bm{x}_1 | \hat \rho (0) | \bm{x}_2 \rangle 
\langle\bm{x}_2| U^\dagger (t) |\bm{k}\rangle  \ , \nonumber \\
& = &  \lim_{t \to \infty} \int d\bm{x}_1 d\bm{x}_2 
\varphi_{\bm{k}} (\bm{x}_1, t) \rho_0 (\bm{x}_1, \bm{x}_2 ) \varphi^*_{\bm{k}} (\bm{x}_2, t) 
\label{eq:P1a}
\\
P_2(\bm{k}_1,\bm{k}_2) 
& = & \lim_{t \to \infty} 
\int \!\!d\bm{x}_1d\bm{x}_2 d\bm{x}^\prime_1 d\bm{x}^\prime_2
\langle \bm{k}_1\bm{k}_2 |U (t) |\bm{x}_1\bm{x}_1^\prime\rangle 
\nonumber \\
& & \qquad \qquad \qquad  \times
\langle \bm{x}_1,\bm{x}_2 | \hat \rho (0) | \bm{x}'_1,\bm{x}^\prime_2 \rangle 
\langle \bm{x}_2\bm{x}^\prime_2|U^\dagger (t) |\bm{k}_1\bm{k}_2\rangle \nonumber \\
& = & \lim_{t \to \infty} \int d\bm{x_1} d\bm{x_2}  d\bm{x}^\prime_1 d\bm{x}^\prime_2
\Psi_{\bm{k}_1,\bm{k}_2}(\bm{x}_1,\bm{x}_2; t)  \rho_0 (\bm{x_1},\bm{x_2}; \bm{x}'_1,\bm{x}'_2 )
\Psi^*_{\bm{k}_1,\bm{k}_2}(\bm{x}'_1,\bm{x}'_2, t) 
 \nonumber \\
\label{eq:P2a} 
\end{eqnarray} 
where 
$\varphi_{\bm{k}} (\bm{x}, t) =  \langle\bm{k}| U(t) |\bm{x}\rangle$
is the amplitude that a particle emitted at $\bm{x}$ is detected with momentum $\bm{k}$ 
at time $t$,
while 
$
\Psi_{\bm{k}_1,\bm{k}_2}(\bm{x}_1,\bm{x}_2, t ) 
=  \langle \bm{k}_1\bm{k}_2 |U (t) |\bm{x}_1\bm{x}_2 \rangle
$
is the two-particle amplitude that two particles are emitted at $\bm{x}_1$ and $\bm{x}_2$ 
and detected with momenta $\bm{k}_1$ and $\bm{k}_2$ at time $t$ simultaneously. 
We have introduced the one-particle reduced density matrix, 
$\rho_1 (\bm{x}_1, \bm{x}_2 ) = \langle \bm{x}_1 | \hat \rho (0) | \bm{x}_2 \rangle$,
and the two particle reduced density matrix, 
$\rho_2 (\bm{x}_1,\bm{x}_2; \bm{x}'_1,\bm{x}'_2 ) 
= \langle \bm{x}_1,\bm{x}_2 | \hat \rho (0) | \bm{x}'_1,\bm{x}^\prime_2 \rangle $
which obey, due to the bosonic symmetry of the two particle states ($| \bm{x}_1,\bm{x}_2 \rangle
= | \bm{x}_2,\bm{x}_1\rangle$) , 
\begin{eqnarray}
\rho_2 (\bm{x}_1,\bm{x}_2; \bm{x}'_1,\bm{x}'_2 ) 
 =  \rho_2 (\bm{x}_2,\bm{x}_1; \bm{x}'_1,\bm{x}'_2 ) = \rho_2 (\bm{x}_1,\bm{x}_2; \bm{x}'_2,\bm{x}'_1 )
\label{rho2}
\end{eqnarray}
The basic objective of the HBT interferometry is to extract information of the density
matrix from the observed correlation function $C(\bm{k}_1,\bm{k}_2)$.  

If the particles do not interact after $t=0$, the time evolution operator turns into a trivial phase 
factor and the single particle amplitude becomes just a plane wave:
$\langle\bm{k}|e^{-i\hat{H}t}|\bm{x}\rangle = e^{- iE_k t} \langle\bm{k}|\bm{x}\rangle
= e^{- i (\bm k \cdot \bm x + E_k t)}$.
This results in 
\begin{eqnarray}
P_1(\bm k) &=& \int d\bm{x}_1 d\bm{x}_2 \rho_0 (\bm{x}_1, \bm{x}_2 ) e^{- i \bm{k} \cdot (\bm{x}_1
- \bm{x}_2 )} 
= \int\!\! d\bm{x}  \ f(\bm{x} ,\bm{k} )    \label{eq:P1free}
\end{eqnarray}
where $f(\bm{x},\bm{k} )$ with $\bm{x} = (\bm{x}_1 + \bm{x}_2 )/2$ is the one-body phase space distribution function or the Wigner function at $t = 0$.
Similarly, in the absence of the final state interaction, the symmetrized two particle 
amplitude becomes,
$\langle \bm{k}_1\bm{k}_2 |U (t) |\bm{x}_1\bm{x}_2 \rangle 
 =  \frac{1}{\sqrt{2}} \left[ e^{ - i ( \bm{k}_1 \cdot \bm{x}_1 + \bm{k}_2 \cdot \bm{x}_2 ) }  + 
e^{ - i( \bm{k}_1 \cdot \bm{x}_2 + \bm{k}_2 \cdot \bm{x}_1 ) }  \right]  e^{- i ( E_1 + E_2)t} $.
If we insert this expression into Eq (\ref{eq:P2a}) 
and further assume the factorization of the two particle density matrix, 
$\rho_0 (\bm{x_1}, \bm{x_2}; \bm{x}'_1,\bm{x}'_2 ) 
= \frac{1}{2} \left[ \rho_0 ( \bm{x}_1, \bm{x}'_1 )  \rho_0 ( \bm{x}_2, \bm{x}'_2 ) + 
 \rho_0 ( \bm{x}_2, \bm{x}'_1 )  \rho_0 ( \bm{x}_1, \bm{x}'_2 )  \right] $,
along with the symmetry (\ref{rho2}), we find
\begin{eqnarray}
P_2(\bm{k}_1,\bm{k}_2) 
= P_1 (\bm{k}_1) P_1 (\bm{k}_2)
+  \left| \int d\bm{x}  f ( \bm{x}, \bar{\bm{k}}) e^{i \bm{q} \cdot \bm{x} } \right|^2 
\label{eq:P2free}
\end{eqnarray}
with $\bar{\bm{k}} = \frac{1}{2} (\bm{k}_1+ \bm{k}_2)$ and 
$\bm{q} = \bm{k}_1 - \bm{k}_2$.
For a factorized form of the phase space distribution 
$f ( \bm{x}, \bm{k} ) = \tilde{\rho} (\bm{x}) P_1 (\bm{k}) $ with 
$\int d \bm{x} \tilde{\rho} (\bm{x}) =1$, 
this gives the familiar form of the correlation function,
 \begin{equation} 
C(\bm{q}, \bar{\bm{k}} ) = 1+  \eta (\bm{k}_1, \bm{k}_2) \left| \int d\bm{x}  \tilde{\rho} ( \bm{x} ) e^{i \bm{q} \cdot \bm{x} } \right|^2 ,
\label{eq:Cphi0}
\end{equation}
with a correction factor 
$\eta (\bm{k}_1, \bm{k}_2) = P_1^2 (\bar{\bm{k}}) /P_1 (\bm{k}_1) P_1 (\bm{k}_2)$, 
which exhibits limiting behaviors: 
$C (0, \bar{\bm{k}}) = 2$ and $C (\infty, \bar{\bm{k}}) = 1$.
This formula has been used to reconstruct the source distribution function 
from the measured two particle momentum correlation.
We note that the lesson we learn from this simple exercise is that the phase 
interference of the symmetrized two-particle amplitude plays essential 
role to create the two particle momentum correlation. 

We now consider how this result is modified in the presence of the final state 
interaction.   We shall study the change of the single particle amplitude 
$
\langle \bk | U(t) | \bx \rangle$ 
as well as the two particle amplitude 
$\langle \bm{k}_1\bm{k}_2 | U(t) |\bm{x}_1\bm{x}_2 \rangle$
by the action of the time evolution operator 
$U(t) = e^{- i \hat{H} t}$.  Here we replace the many-body Hamiltonian $\hat H$ 
by one-body Hamiltonian for a non-relativistic particle propagating in an 
one-body potential. 
Without mutual two-body interaction of pion pair, 
two-body amplitude 
in (\ref{eq:P2a}) is expressed by the symmetrized products of one-body amplitudes :
\begin{eqnarray}
\langle \bm{k}_1\bm{k}_2 |U (t) |\bm{x}_1\bm{x}_2 \rangle = 
\frac{1}{2} \left\{ \langle\bm{k}_1| U(t) |\bm{x}_1\rangle\langle\bm{k}_2| U(t) |\bm{x}_2\rangle +
\langle\bm{k}_1| U(t) |\bm{x}_2\rangle\langle\bm{k}_2| U(t) |\bm{x}_1\rangle\right\}
  \label{eq:2amp}
\end{eqnarray}
We therefore only need to  compute the distortion of the single particle amplitude. 

We recall that 
$\varphi_\bk (\bx, t ) =  \langle \bk | U(t) | \bx \rangle$ 
denotes the amplitude for a particle emitted at $\bx$ at time $t=0$
being detected with asymptotic momentum $\bk$ at sufficiently large time $t$.   
Formally, it may be interpreted as a wave function which obeys the static 
one-body Sch\"ordinger equation on the transverse plane:  
\begin{eqnarray} 
\left[ -\frac{1}{2m} \nabla^2 + V(\bx ) \right] \varphi_\bk (\bx, t) 
= E_\bk \varphi_\bk (\bx, t )
\label{eq:Schrodinger}
\end{eqnarray}
with $E_\bk = \bk^2 / 2 m$. 
If we assume cylindrical symmetry for the collision volume, this Schr\"odinger 
equation in two space dimension may be reduced to a one-dimensional one 
in a "central" potential $V(|\bx| )$ and may be solved by the standard WKB 
semiclassical approximation. 

Here we adopt more intuitive approach to the problem motivated by 
the stationary phase approximation of the path integral expression 
of the amplitude, exploiting an analogy to geometrical optics. 
We write
\begin{equation}
\langle \bk | U(T) | \bx \rangle = \int d \bx'  e^{-i \bk \cdot \bx' } \langle \bx' | U (T) | \bx \rangle
\label{eq: 1amp}
\end{equation}
and introduce an Ansatz 
\begin{eqnarray}
 \langle \bx' | U (T) | \bx \rangle  \approx  A (\bx', \bx ; T ) e^{i S_c (\bx', T; \bx , 0 ) }
\label{eq:1amp}
\end{eqnarray}
for the propagator.
We work with the classical action integral
\begin{equation}
S_c (\bx_2, t_2 ; \bx_1, t_1 ) = \int_{t_1}^{t_2} 
d t  L ( \bx, \dot{\bx} ) 
\end{equation}
along the classical trajectory $( \bx (t) , \bp (t) )$ specified by the initial 
position $\bx_1$ of the particle at time $t = t_1$ and the final position 
$\bx_2$ at time $t = t_2$.  
The initial and final momenta of the particle are given by
$\bp_1 =  - \nabla_{1} S_c (\bx_2, t_2 ; \bx_1, t_1 )$ 
and $\bp_2 =  \nabla_{2} S_c (\bx_2, t_2 ; \bx_1, t_1 ) $
with the nablas $\nabla_{1,2}$  operating on $\bx_{1,2}$, respectively. 
For free particle motion
\begin{equation}
S_c^{\rm free} (\bx', T ; \bx, 0) =  \frac{m(\bx' - \bx )^2}{2T}
\end{equation}
so that the integration over $\bx'$, assuming that the prefactor 
$A(\bx,\bx', T)$ is a constant, evidently give the previous results:
$
\langle \bk | U(T) | \bx \rangle = e^{- i \bk \cdot \bx - i E_\bk T }
$. 
This motivates us to write:
\begin{equation}
\langle \bk | U (T) | \bx \rangle = {\cal A} (\bx, \bk ) e^{i W (\bx, \bk ) - i E_\bk T} 
= \varphi_\bk (\bx) e^{- i E_\bk T} 
\label{eq: W}
\end{equation}
where we separated the phase factor into an explicitly time-dependent 
piece $E_\bk T$ and $T$-independent part $W (\bx, \bk )$.    Evidently,
for free particle, $W (\bx, \bk ) = W_0 (\bx, \bk ) = - \bk \cdot \bx$. 

To construct the phase factor $W (\bx, \bk )$ in the presence
of interaction, we decompose the classical action into two parts:
\begin{equation*}
S_c (\bx', T; \bx , 0 ) = S_c ( \bx', T; \bx'', T' ) + S_c ( \bx'', T'; \bx, 0 )
\end{equation*}
where $(\bx'', T' )$ is chosen at any point $\bx'' (T')$ on the classical trajectory.   
If we choose a point sufficient far away from the interacting 
region, we may set
\begin{equation*}
S_c ( \bx', T; \bx'', T' ) = S_0  ( \bx', T; \bx'', T' ) = \frac{m(\bx' - \bx'')^2}{2 (T-T')}
\end{equation*}
Putting this into (\ref{eq:1amp}) and performing the integral over $\bx'$, 
we obtain,
\begin{equation}
\langle \bk | U (T) | \bx \rangle = A (\bx', \bx, T ) 
e^{- i E_\bk ( T - T' ) - i \bk \cdot \bx'' + i S_c ( \bx'', T'; \bx, 0) }
\end{equation}
Comparison with (\ref{eq: W}) gives the formula:
\begin{equation}
W (\bx, \bk ) = E_\bk T'  -  \bk \cdot \bx'' +  S_c ( \bx'', T'; \bx, 0) 
\end{equation}
and ${\cal A} (\bx, \bk ) = A (\bx', \bx ; T )$.
We emphasize that $(\bx'', T')$ is just a dummy coordinate which can be arbitrarily 
chosen on the classical trajectory as long as $T'$ is taken to be sufficiently large
so that $W$ is a function only of $\bx$ and $\bk$.

This procedure determines $W (\bx, \bk)$ uniquely if the classical trajectory 
corresponding to $(\bx,\bk)$ is given. 
There are cases, however, that there are more than one classical trajectories 
which correspond to the same $\bx$ and $\bk$, 
similar to appearance of caustics in geometrical optics although time is reserved 
in our case.  
We will see this actually occurs in the presence of interaction.

In the presence of the imaginary part $V_2 (\bx) $ of the optical potential the prefactor
may be given 
\begin{equation}
{\cal A} (\bx, \bk ) = a (\bk ) \exp \left[ \int_0^\infty dt V_2 (\bx (t)) \right]
\label{eq:prefactor}
\end{equation}
where the integration is along the classical trajectory determined by the real
part of the optical potential $V_1 (\bx)$ for given $\bx$ and $\bk$.
This integral is negative for absorptive potential $V_2 (\bx) < 0$. 

We first consider how $P_1 (\bk)$ is modified by the interaction.
Using the center-of-mass coordinate $\bm{x} = \frac{1}{2} ( \bx_1 + \bx_2 )$ and 
relative coordinate $\bm{r} = \bx_1 - \bx_2$, we have
$\varphi_\bk (\bx_1)  =  \varphi_\bk (\bx + \frac{1}{2} \br )  
\simeq \exp \left[ i \frac{1}{2} \bm{r} \cdot \nabla W (\bx, \bk ) \right]  \varphi_\bk (\bx )$,
and 
$\varphi_\bk (\bx_2)  =  \varphi_\bk (\bx - \frac{1}{2} \br ) 
\simeq \exp \left[ - i \frac{1}{2} \bm{r} \cdot \nabla W (\bx, \bk ) \right]  \varphi_\bk (\bx )$,
ignoring the higher order derivatives of  $W (\bx, \bk)$ with respect to $\bx$,
as well as the derivatives of the prefactor $A(\bx, \bk)$ with respect to $\bx$.  
This procedure may be justified for our choice of $V_1( \bx)$ which is a smooth function 
in the source region.    
Inserting the above results in (\ref{eq:P1a}) we find
\begin{eqnarray}
P_1(\bm{k}) 
& = & \int d\bm{x} d\bm{r}  \int \frac{d \bk}{(2 \pi)^2}  e^{ - i \br \cdot (\bk - \nabla W (\bx, \bk) )} f (\bx, \bk ) 
A^2(\bx, \bk ) 
\nonumber \\
& = & a^2 (\bk) \int d \bx f (\bx, \bp (\bx, \bk)) e^{  - 2 \gamma (\bx, \bk )} 
\end{eqnarray}
where $\bp (\bx, \bk ) = \nabla W ( \bx, \bk ) $ is the initial momentum of the particle
when it is emitted and $\gamma (\bx, \bk ) = -  \int_0^\infty  dt V_2$ measures the 
amount of absorption.  

This result is precisely what is expected by our intuition.  
The real part of the optical potential causes the shift of the momentum of the observed 
particle from the original momentum at the time of emission by acceleration by the
potential field.  This is purely classical effect which may be obtained from the classical
treatment\cite{BBM96}. 
In the quantum description, the information of the momentum shift of the particle 
is encoded in the phase shift of the single particle amplitude in proportion to $\bx$.  
The flux attenuation factor $e^{  - 2 \gamma (\bx, \bk )}$ is due to the absorption (including
elastic scattering of the particle) on the way out of the emission point which may also
included on purely classical consideration. 

We now study how this phase shift results in the change of the two-particle momentum 
correlation through the change of interference which arises from the symmetrization of
the two-particle amplitude.
The two particle joint probability (\ref{eq:P2a}) may be evaluated with the same technique.
Inserting
\begin{eqnarray*}
\Psi_{\bm{k}_1,\bm{k}_2}(\bm{x}_1,\bm{x}_2, t)  = 
\frac{1}{\sqrt{2}} \left[ \varphi_{\bk_1} (\bx_1) \varphi_{\bk_2} (\bx_2) 
+ \varphi_{\bk_2} (\bx_1) \varphi_{\bk_1} (\bx_2) \right] e^{ - i (E_1 + E_2 ) t} 
\end{eqnarray*}
and similar form for $\Psi^*_{\bm{k}_1,\bm{k}_2}(\bm{x}'_1,\bm{x}'_2, t) $ in (\ref{eq:P2a}), 
and then assuming again the factorization of two-body reduced density matrix, 
we obtain
\begin{eqnarray}
P_2 ( \bk_1, \bk_2 ) = P_1 (\bk_1) P_1 (\bk_2) + 
\left| F ( \bk_1, \bk_2 ) \right|^2 
\end{eqnarray}
where we have introduced the integral
 \begin{equation}
 F ( \bk_1, \bk_2 ) =  \int d \bx_1 d \bx_2  \rho_1 ( \bx_1, \bx_2)
\varphi_{\bk_1} (\bx_1) \varphi^*_{\bk_2} (\bx_2) ,
\end{equation}
to express the interference term.
Using the Ansatz (\ref{eq:1amp}) and (\ref{eq:prefactor}), 
\begin{eqnarray}
F ( \bk_1, \bk_2 ) & = & a (\bk_1) a (\bk_2) 
\int d \bx_1 d \bx_2  \rho ( \bx_1, \bx_2) 
 e^{  i ( W (\bx_1, \bk_2 ) - W (\bx_2, \bk_1 ) ) }  e^{ -  \gamma (\bx_1, \bk_2) - \gamma (\bx_2, \bk_1) }
 \nonumber \\
\end{eqnarray}
We approximate the exponents by
$W (\bx_1, \bk_1 ) - W (\bx_2, \bk_2 )  =  W (\bx + \frac{1}{2} \bm{r} , \bk +  \frac{1}{2} \bm{q} ) 
- W  (\bx - \frac{1}{2} \bm{r} , \bk - \frac{1}{2} \bm{q} ) =
 \simeq  \bm{r} \cdot \nabla W (\bx, \bk ) + \bm{q} \cdot \nabla_\bk W (\bx, \bk ) $ and 
$ \gamma (\bx_1, \bk_2) + \gamma (\bx_2, \bk_1)  \simeq  2 \gamma (\bx, \bk )$ ,
retaining only the leading terms in the expansion in $\bm{q}$ 
since we expect to find a significant momentum correlation only for small $q$ less
than the inverse of the linear dimension of the source. 
Inserting $\rho ( \bx_1, \bx_2)  = \frac{1}{(2\pi)^2} \int d \bp f (\bx, \bp ) e^{-i \bp \cdot \bm{r} }$
and then performing integral over $\bm{r}$, we find
\begin{eqnarray}
F ( \bk_1, \bk_2 ) 
& = &  a (\bk_1) a (\bk_2) \int d \bx  f ( \bx, \bp (\bx, \bk ) ) 
 e^{  - i \bm{q} \cdot \nabla_\bk W (\bx, \bk ) } e^{ - 2 \gamma (\bx, \bk) } 
\label{eq:F}
\end{eqnarray}
where $\bp (\bx, \bk ) = \nabla W (\bx, \bk )$ is the initial momentum of the particle.

For free particles, we may set $W (\bx, \bk ) = W_0 (\bx, \bk) = - \bx \cdot \bk$, 
$a (\bk) =1$, $2 \gamma (\bx, \bk ) =0$ and $\bk = \bp$ so that (\ref{eq:F}) is reduced to
\begin{eqnarray}
F ( \bk_1, \bk_2 ) =  \int d \bx  f ( \bx, \bk )  e^{ i \bm{q} \cdot \bx  } 
\end{eqnarray}
which reproduces (\ref{eq:P2free}).

To see how the interaction modifies the source image we write
\begin{equation}
W (\bx, \bk ) = W_0 (\bx, \bk) + \delta W (\bx, \bk) = - \bx \cdot \bk +  \delta W (\bx, \bk) 
\end{equation}
where $\delta W (\bx, \bk )$ is the phase shift by the mean field interaction. 
Since $\nabla_\bk W (\bx, \bk ) = - \bx + \nabla_\bk \delta W (\bx, \bk) $, this gives
 (\ref{eq:F}) as 
\begin{equation}
F ( \bk_1, \bk_2 ) = a (\bk_1) a (\bk_2) \int d \bx  f ( \bx, \bp (\bx, \bk) ) 
 e^{  i \bm{q} \cdot ( \bx - \nabla_\bk \delta W (\bx, \bk ) ) } e^{ - 2 \gamma (\bx, \bk) } 
\end{equation}
Note that the phase shift $\delta W (\bx, \bk)$ caused by the real part of the mean field
potential resulted in the apparent shift of the emission point by 
$\delta \bx = - \nabla_\bk \delta W (\bx, \bk )$
while the imaginary part effectively cut off the contribution of the source from 
deep interior and the back side.
Transforming the integration variable $\bx$ to 
\begin{equation}
\bx' = \bx - \nabla_\bk \delta W (\bx, \bk )
\label{eq: x2x'}
\end{equation}
we can write 
\begin{equation}
F ( \bk_1, \bk_2 ) = a (\bk_1) a (\bk_2) \int d \bx'  f_{\rm eff} ( \bx', \bk )  
 e^{  i \bm{q} \cdot \bx'  } 
\end{equation}
with
\begin{equation}
f_{\rm eff} ( \bx', \bk ) =  J ( \bx, \bx'; \bk ) f ( \bx, \bp (\bx, \bk) ) e^{ - 2 \gamma (\bx, \bk) } 
\label{eq:f_eff}
\end{equation}
where $J (\bx, \bx'; \bk ) = \partial ( \bx, \bk )  / \partial ( \bx', \bk) 
= [ \partial ( \bx', \bk) / \partial ( \bx, \bk ) ]^{-1}$ is the Jacobian of the coordinate 
transformation $( \bx, \bk)  \to ( \bx', \bk)$ defined by (\ref{eq: x2x'}) and
$\bx$ on the right-hand side is understood as a function of $\bx'$ and $\bk$. 
The correlation function is now given by
\begin{eqnarray}
C (\bk_1, \bk_2)  &=& 1+ \left| \int d\bx'
\tilde \rho_{\rm{eff}} ( \bx' , \bk) e^{i \bm{q} \cdot \bx' } \right|^2 
\end{eqnarray}
with the effective source distribution defined by
\begin{equation}
\tilde \rho_{\rm{eff}} ( \bx', \bm{k})  =  
\frac{ f_{\rm eff} ( \bx', \bk ) }
{\int d\bx f ( \bx, \bm{p} (\bx,\bk)) e^{-  2 \gamma ( \bx, \bk ) } }
 \label{eq:rho_eff}
\end{equation}
Note that the momentum dependent prefactors $a(\bk_1) a (\bk_2) $ cancel out 
by the normalization.   The absorption factor $e^{-2\gamma (\bx, \bk)}$ gives a 
weight on the distribution of the emission points toward the side of the 
observation of the particle.   

For numerical computation we have chosen here a schematic pion optical 
potential with two-range Gaussian shape for the real part, 
$V_r ( \bx  ) = {\cal V}_1 e^{- \bx^2 / 2 \lambda_1^2} 
+ {\cal V}_2 e^{- (|\bx| - \xi)^2/ 2 \lambda_2^2} $, 
and one-range Gaussian for the imaginary part, 
$V_i ( \bx ) =  - {\cal V}_i e^{- \bx^2  / 2 \lambda_1^2} $,
where the shorter range  $\lambda_1$ of the potentials is taken to be the same as the 
parameter for the Gaussian source distribution,  
$\rho ( \bx ) = \rho_0 e^{-  \bx^2  / 2 \lambda_1^2} $, 
while $\lambda_2$ gives the thickness of the meson hallow surrounding 
the dense meson source.
We have chosen $\lambda_1 = 5$ fm, $\lambda_2 = 5$ fm, $\xi = 10$ fm, 
${\cal V}_1 = 10$ MeV, ${\cal V}_2 = 2$ MeV, ${\cal V}_i = 0.1$ MeV. 
We also used thermal Bose-Einstein distribution of the initial pion momentum 
with temperature $T_0 = 140$MeV.    

The classical trajectories are
computed numerically with the initial condition $(\bx, \bp)$ until the asymptotic 
momentum $\bk $ is reached.  Since the total energy is conserved the magnitude 
of $\bk $ is determined by the initial value of $|\bp|$ and $V_1 (\bx)$.   
The integration gives the angle of deflection with respect to the initial 
direction of emission $\hat \bp$. 
Along each trajectory the action integral is computed numerically; this 
determines the phase factor $W (\bx, \bk)$ as a function of $(\bx, \bk)$. 
Computing many such trajectories with different initial conditions, 
we have a good sample of trajectories and the phase factor $W (\bx, \bk)$.
We use these data set to compute the derivatives of $\delta W (\bx, \bk)$ 
with respect to $| \bk |$ and $\theta_\bk$ by interpolation which are
related to the apparent location of the emission points in Cartesian 
coordinate, $ \bx' = ( x', y' )$, where 
$x'  = x - \frac{\partial \delta W}{\partial  \bar{\bm k}}$ is the apparent location 
of the emission point in the direction of the final meson momentum $\bk$ 
(outward direction), while 
$y' = y -  \frac{1}{\bar{k}} \frac{\partial \delta W}{\partial \theta_{\bk}}$ is 
the apparent location perpendicular to $\bk$ (sideward direction) where 
$\theta_\bk$ is the angle of $\bk$.

 \begin{figure}
  \begin{center}
      \includegraphics[scale=0.5]{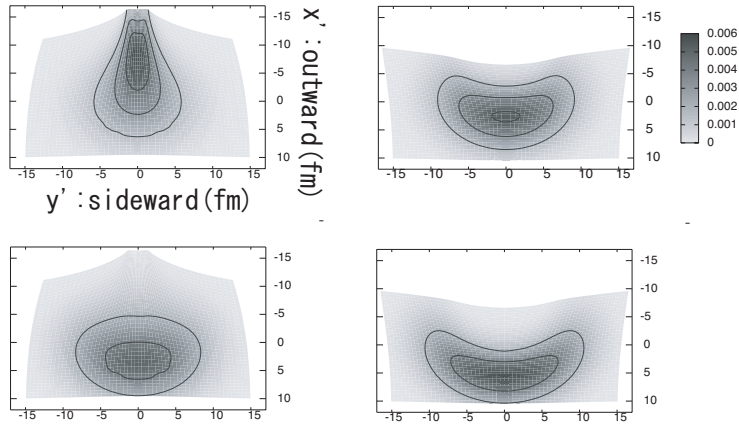}
  \end{center}
   \caption{Contour plots of the effective source distribution $\tilde{\rho}_{\rm eff.} (\bx', \bk)$ 
 defined by (31) at $k=100$MeV/c.  Left (right) columns correspond to results with 
 repulsive (attractive) potential.  Lower panels include effect of absorption. 
 Vertical (horizontal) coordinate denotes the shifted outward (sideward) location of
 the emission points whose plot range is distorted by the non-linear coordinate transformation. }
   \label{fig:cntr1}
 \end{figure}
 
  \begin{figure}
  \begin{center}
        \includegraphics[scale=0.5]{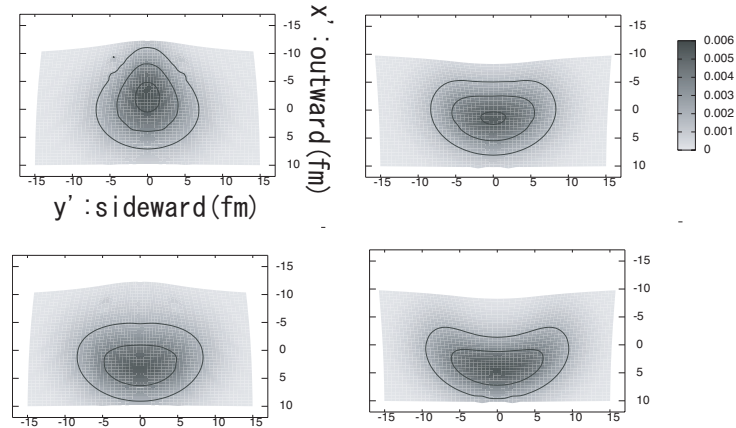}
  \end{center}
   \caption{Contour plots of the effective source distribution at $k=150$MeV/c.  Each panel
   corresponds to the same choice of the potential at the same location in Fig.1}
   \label{fig:cntr150}
 \end{figure}

We show in Fig. 1 contour plots of the effective source image at $k=100$MeV/c. 
The upper two panels give the results with no absorption (${\cal V}_i = 0$)
while the lower panels are computed with ${\cal V}_i =  0.1$ MeV. 
The vertical and  horizontal coordinates gives the apparent location of the 
emission points in the outward direction (direction of $\bk$) and the 
sideward direction (direction perpendicular to $\bk$) respectively.  
We first compare the case for repulsive interaction ${\cal V}_1 = 10$ MeV 
with no absorption  (upper left panel) and the attractive interaction with no 
absorption (upper right panel).  
We found that the real part of potential affects on the extension of source 
in both outward and sideward directions.  The repulsive interaction 
leads to elongation of the effective source image in the outward direction, 
while the attraction tends to shrink the source in this direction.  
On the other hand, the repulsive force leads to shrinking sideward source 
extension, while the attractive force gives the sideward extension 
stretched.   These effects are seen as a geometrical effect of either 
stretching or compressing the scale of each coordinate.   The distortion
of the original image is weakened as the momenta of the two
particles increases as seen in Fig. 2 at $k = 150$MeV/c.

Our results qualitatively agree with the results of two other groups\cite{CMWY,Pratt06}. 
Pratt and Cramer et al. suggested that this change of the apparent source 
size in the sideward direction may be interpreted as due to the refraction 
or the lensing effect in geometrical optics.   In our analyses, however, 
the apparent shift of the emission point is caused by the $\bk$ dependence 
of the phase shift $\delta W (\bx, \bk)$, through Eq. (26): the apparent shift of 
the emission point $\delta \bx = - \nabla_\bk \delta W (\bx, \bk )$ arises due to 
the difference of the action of the potential for particles traversing along two 
adjacent trajectories starting at the same point $\bx$ ending up with different 
final momenta.  

To make this issue more quantitative, we examine 
the phase shift using Glauber-type approximation which assumes 
the straight-line trajectories in the interaction regions so that the phase shift is 
given by the simple formula which may be written for $V << \bk^2/2m$,  
\begin{eqnarray}
\delta W_{\rm Glauber} (\bm x,\bk) 
&\simeq& - \frac{m}{|\bk|} \int_{u (0)}^{u (T)} V(\sqrt{u^2 + b^2}) du \label{eq:SGlauber}
\end{eqnarray}
where $b = \sqrt{ \bx^2 - (\bx \cdot \bk)^2/ \bk^2}$ is the "impact parameter" of the trajectory, 
and $u (t) = \bx(t) \cdot \bk/ |\bk|$ is the position of the particle along the trajectory 
at time $t$.  This phase shift depends on $\theta_\bk$ since $b$ is related to 
$\bk$ by $b =  \sqrt{ \bx(T)^2 - (\bx(T) \cdot \bk)^2/ \bk^2}$ at large time $T$.
The change of the HBT image calculated by this formula is shown in Fig.3 
for the same cases as in Fig.1.   It is seen that the "Glauber approximation" 
qualitatively reproduces the same results.   This implies that the deflection of the
classical trajectory in the source region is not essential for the distortion of the
HBT images; rather, the change of the relative momentum of two particles via the
difference of their phase shifts is the origin of the distortion of the source image.  

\begin{figure}
  \begin{center}
        \includegraphics[scale=0.5]{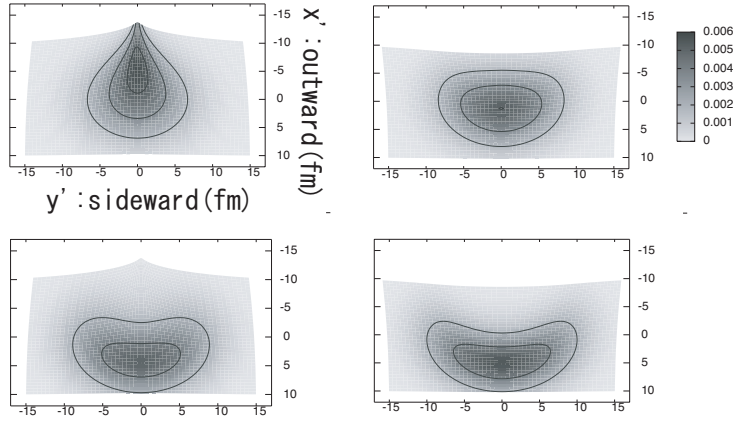}      
  \end{center}
  \caption{The effective source distributions at $k=100$MeV/c in the Glauber approximation 
  computed with the same potentials as in Fig.1}
   \label{fig:Glauber}
 \end{figure}

The inclusion of the absorption, however, diminishes all these interesting
effects of the mean field interaction, by effectively cutting off the contributions 
from deep interior of the source as well as the other side of the source.  
The result is somewhat similar to the effect caused by the attractive interaction, 
namely the source image is effectively stretched in the sideward direction. 

Although we still need to make improvements in our calculations (relativistic
treatment, time-dependent source structure\cite{BB89,RG96}, more realistic pion 
optical potential, etc.) before confronting the experimental data, 
these mean field effects may play some role in reducing the discrepancy between 
the data and hydrodynamic simulations at small values of $k$.

In conclusion, we have studied the effect of the final state interaction in
the meson clouds on the HBT interferometry in heavy-ion collisions and
have shown that the final state interaction causes a significant distortion of the
source images at small $k$ through the change of the phase shift in the single 
particle amplitude and the absorption effect.   
More detail account of this work will be reported elsewhere. 

\section*{Acknowledgements}
We thank Gordon Baym, Tetsufumi Hirano and Koichi Yazaki for helpful 
conversations on the related works, and Hirotsugu Fujii for calling our attention 
to the reference \cite{Pratt06}.
This work is supported in part by the Grants-in-Aid of MEXT, Japan,  
No. 19540269, and 
Global COE Program "the Physical Sciences Frontier", MEXT, Japan. 

%

\end{document}